\documentclass{article}
\usepackage{epsfig}

\title{\bf Nuclear Matter in Intense Magnetic Field and Weak 
Processes }

\author{Ashok Goyal\thanks{E--mail:agoyal@ducos.ernet.in},
V.K.Gupta,Kanupriya Goswami and Vinita Tuli\\
{\em  Department of Physics and Astrophysics,}\\
{\em  University of Delhi,Delhi-110007,India.}\\
{\em  Inter University Centre for Astronomy and Astrophysics}\\
{\em  Ganeshkhind,Pune-411007,India.}\\
}

\begin{document}

\maketitle

\begin{abstract}
\noindent We study the effect of magnetic field on the dominant
neutrino emission processes in neutron stars.The processes are first
calculated for the case when the magnetic field does not exceed the
critical value to confine electrons to the lowest Landau state.We then
consider the more important case of intense magnetic field to evaluate
the direct URCA and the neutronisation processes. In order to estimate
the effect we derive the composition of cold nuclear matter at high
densities and in beta equilibrium, a situation appropriate for neutron
stars. The hadronic interactions are incorporated through the exchange
of scalar and vector mesons in the frame work of relativistic mean
field theory. In addition the effects of anomalous magnetic moments of
nucleons are also considered.    
\end{abstract}
\pagebreak

\begin{section} {Introduction}
\indent Recently there has been a lot of interest in the study of the
effects of strong magnetic fields on various astrophysical
phenomena.Many of these require large magnetic fields and indeed fields 
 with a strength
of $10^{18}$ Gauss or even more,have been conceived to exist at the time
of the supernova collapse inside neutron stars and in other
astrophysical compact objects {\cite{label 1}}.The matter density in neutron 
stars
exceeds the nuclear density and  the  presence of high magnetic fields
modifies the energy of charged particles confining them to low landau
levels.This results in shifting of beta equilibrium and the
composition of matter gets significantly modified 
{\cite{label 2},\cite{label 3}}.The intense magnetic fields also result in 
reduced photon opacities and enhanced cooling of neutron stars 
{\cite{label 4}} .Effect of magnetic field on weak interaction rates is of 
great importance in neutrino emission in neutron stars and during first 
few seconds immediately after collapse of massive neutron stars 
{\cite {label 5}},and also in light elements abundances in the early universe
{ \cite{label 6}}.  Specifically it has been 
shown { \cite{label 7}} that intense magnetic field induces an asymmetry in 
neutrino emission which may be capable of imparting a large recoil momentum 
to the newly born neutron star.
The large magnetic field would also affect the location of electron neutrino 
sphere through modification of neutrino scattering cross-section thus 
affecting the Kasenko-Segre{\cite{label 8}} mechanism to account for pulsar 
picks.      
The dominant mode of energy loss in neutron stars is through neutrino emission.
The important process leading to neutrino cooling are the so called 
URCA

\vspace{0.5cm}
\begin{equation}
n\longrightarrow p+e^{-}+\bar{\nu_{e}} 
\end{equation}

\begin{equation}  
p+e^{-}+\longrightarrow n+\bar{\nu_{e}}
\end{equation}
 
modified URCA

\begin{equation}
N+N \rightarrow N+N+\nu+\bar{\nu} 
\end{equation}

\begin{equation}		
n+n\rightarrow n+p+e^{-}+\bar{\nu_{e}}
\end{equation}

\begin{equation}	
n+p+e^{-}\rightarrow n+n+\nu_{e}
\end{equation}

and the pion mediated
\begin{equation}
n+\pi^{-}\longrightarrow n+e^{-}+\bar{\nu_{e}}
\end{equation}
process accompanied by along with their inverse reactions.

\noindent At low temperatures for degenerate nuclear matter, the direct URCA
process can take place only near the fermi energies of participating particles and
simultaneous conservation of energy and momentum require the inequality
$p_{F}(e)+p_{F}(p)\geq p_{F}(n)$ to be satisfied in the absence of the 
magnetic field.This leads
to the well known threshold {\cite{label 9}} for the proton fraction 
$Y_{p}=\frac{n_{p}}{n_{B}} \geq 11 \% $ thus leading to strong suppression in 
nuclear matter.This condition is satisfied for $n_{b} \geq 1.5 n_{0}$( where $n_{0}=0.16 fm^{-3}$,is the nuclear saturation density),in a relativistic
mean field model of interacting n-p-e gas for $B=0$.
In this paper we calculate the energy loss due to the above process firstly
for the case when the magnetic field does not exceed the critical value to
confine degenerate electrons to the lowest Landau level. The interesting
case arises when the magnetic field is intense enough to force electrons into
the ground state such that electrons occupy the lowest Landau level with their
spins aligned opposite to the magnetic field direction.the charge neutrality
now pushes the protons too into the ground state.In this situation we require
to use the exact wave functions obtained by solving Dirac equation in the
magnetic field. We calculate the energy loss rate due to the direct URCA
process which now proceeds regardless of the proton fraction. We also 
calculate the neutronization rate

\begin{equation}
e^{-} + p \rightarrow n + \nu_{e}
\end{equation}

for a situation appropriate during collapse.

\noindent In order to make numerical estimates of the cooling rates in 
magnetised
neutron stars,we require the composition of nuclear matter.For this we
consider electrically neutral nuclear matter composed of  relativistic
nucleons and electrons in beta equilibrium.We take the hadronic interactions through 
$(\rho-\omega-\sigma)$ meson exchange in the framework of relativistic nuclear mean field theory as in {\cite{label 3}}.In addition we also
incorporate the anomalous magnetic moments of nucleons.Indeed for magnetic
fields strong enough to shift the masses due to magnetic moment nearly equal
to the magnitude of nucleon masses themselves, we may not be able to define
the magnetic moments and we need a field theoretical calculation of self
energy of particles in the external magnetic field.Such a calculation for
 electrons  had been performed by Schwinger{\cite{label 10}} but unfortunately
is not available for nucleons for we only have phenomenological anomalous 
magnetic moments for these particles.When considering very intense magnetic 
fields we should be careful about the effect of magnetic field on strong 
interactions because in this situation,the nucleons and mesons interact both
with  with the magnetic fields through their charges and magnetic 
moments.However for magnetic fields not much greater than $10^{18}$  Gauss we 
do not have to worry about this problem.

In Section II,we obtain the composition of interacting nuclear matter in the
presence of magnetic field by taking into consideration the anomalous
magnetic moments of the nucleons.In section III we calculate the energy
loss due to URCA and modified URCA process for moderate magnetic fields.The 
direct URCA process is then  studied in the presence of quantising magnetic field 
where the energy -momentum conservation inequality on fermion momenta of 
participating particles breaks down and the neutrino emission proceeds at all 
densities.In Section IV,we present the results and discussions.
\end{section}

\begin{section}{Nuclear matter composition in magnetic field}
	
\noindent We consider here the effect of strong uniform magnetic field on the
composition of nuclear matter consisting of neutrons,protons and electrons in
beta equilibrium and the neutrinos/antineutrinos are assumed to freely stream
out of the system.Hadronic interactions are taken into account by considering
the nucleons to interact by exchanging scalar $\sigma$ and vector($\omega,\rho$)
 mesons
in the framework of relativistic nuclear mean field theory.This is taken into
account in the high density limit by replacing the meson fields by their
classical averages in the equations of motion based on a Lagrangian depicting
the physical process.In addition we also consider the case when nucleons
have anomalous magnetic moments.As discussed in the Introduction since we do
not have a proper field theoretical framework for nucleons,we treat them as 
point particles with given phenomenological anomalous magnetic moments and 
restrict ourselves to fields which do not substantially change the energy of 
these particles.
In a uniform magnetic field B along z axis corresponding to the choice of the
gauge field $A^{\mu}=(0,0,xB,0)$,the relativistic mean field Lagrangian can be
written as
\begin{eqnarray}
{\cal L}
&=&
\sum \bar{\psi_{i}}[i\gamma_{\mu}D^{\mu}-m_{i}+g_{\sigma}\sigma
    -g_{\omega}\gamma_{\mu}\omega^{\mu}
    -\frac{1}{2}g_{\rho}\gamma_{\mu}\overrightarrow{\tau}
\cdot \overrightarrow{p^{\mu}}                   
 + \kappa_{i} \sigma_{\mu \nu} F^{\mu \nu}]\psi_{i}  \nonumber \\
&&{} +\frac{1}{2}[(\partial^{\mu}\sigma)^2-(m_{\sigma}^2\sigma^2)] 
 -\sum_{\omega,p} [\frac{1}{4}(\partial_{\mu}V^{i}_{\nu}
	-\partial_{\nu}V^{i}_{\mu})^2
	-\frac{1}{2}m_{vi}^2(V_{\mu}^{i})^2]
\end{eqnarray}
										
in the usual notation with $D^{\mu}=\partial^{\mu}+ieA^{\mu}$ and $K_{i}$ as the
anomalous magnetic moments given by

\begin{equation}
\kappa_{p}=\frac{e}{2m_{p}}[\frac{g_{p}}{2}-1]
\end{equation}

\begin{equation}		
\kappa_{n}=\frac{e}{2m_{n}}\frac{g_{n}}{2}
\end{equation}

where $g_{p}=5.58$   and $g_{n}=-3.82$  are the Lande's g-factor for protons and neutrons respectively.
Writing the general solution as 

\begin{equation}
\psi(r)=e^{-i(Et-p_{y}y-p_{z}z)}f_{s}(x)\nonumber
\end{equation}
where $f_{s}(x)$ is the four component Dirac solution.Replacing the meson
fields in the relativistic mean field approximation by their density dependent
average values $<\sigma_{0}>$,$<\omega_{0}>$ and $<\rho_{0}^3>$ respectively, the 
equation of motion satisfied by the nucleons in the magnetic field become
\vskip 0.5cm
\begin{eqnarray}
&&{}[-i\alpha_{x}\frac{\partial}{\partial_{x}}+\alpha_{y}(p_{y}-eBx) 
+\alpha_{z}p_{z}+\beta(m_{i}+g_{\sigma}<\sigma_{0}>)-i\kappa_{i}
\alpha_{x}\alpha_{y}B]f_{i,s} \nonumber  \\
&&{}= (E^{i}-U_{0}^i)f_{i,s}(x) 
\end{eqnarray}
 
where
\begin{equation}
U_{0}^{p,n}=g_{\omega}<\omega_{0}> \mp \frac{g_{\rho}}{2}<\rho_{0}^3>
\end{equation}

The equations  are first solved for the case when momentum along the magnetic
field direction is zero and then boosting along that direction till the
momentum is $p_{z}$.For the protons and neutrons we thus get

\begin{eqnarray}
{\cal E}_{\nu,s}^{p}
&=& \sqrt{m_{p}^{*2}+p_{z}^2+eB(2\nu+1-s)
	+\kappa_{p}^2 B^2-2\kappa_{p} B s\sqrt{m_{p}^{*2}+eB(2\nu+1-s)}}\nonumber  \\
&=& E_{\nu,s}^{p}-U_{0}^{p}
\end{eqnarray}
and
\begin{eqnarray}
{\cal E}_{s}^{n}
&=& 
\sqrt{m_{n}^{*2}+\overrightarrow{p^{2}}+\kappa_{n}^2 B^2
- 2\kappa_{n} B s\sqrt{p_{x}^{2}+p_{y}^{2}+m_{n}^{*2}}} \nonumber  \\
&{}& = E_{s}^{n}-U_{0}^{n}
\end{eqnarray}
respectively.
In the above $\nu$ denotes the Landau level and takes values 0,1,2.... and 
$s=\pm 1$ indicates whether the spin is along or opposite to the 
direction of the
magnetic field and $m_{i}^{*}=m_{i}+g_{\sigma}<\sigma_{z}>$ is the effective
mass.
The positive energy spinors for protons are then given by
\begin{equation}
 f_{p,1}(x) = C_{\nu,1}e^{- \frac{\xi^2}{2}} \left(
		\begin{array}{c}
		H_{\nu}(\xi)\\
		0\\
		\frac{p_{z}}{E_{\nu,1}^{p}+m_{p}^{*}}H_{\nu}(\xi)\\
		\frac{i\sqrt{eb}}{E_{\nu,1}^{p}+m_{p}^{*}}H_{\nu+1}(\xi)
		\end{array}
		\right)
\end{equation}

and

\begin{equation}
f_{p,-1}(x)=C_{\nu,-1}e^{- \frac{\xi^2}{2}} \left(
	\begin{array}{c}
	0\\
	H_{\nu}(\xi)\\
	\frac{-2{\nu i} \sqrt{eB}}{E_{\nu,-1}^{p}+m_{p}^{*}} H_{\nu - 1}(\xi)\\
	\frac{-p_{z}}{E_{\nu,-1}^{p}+m_{p}^{*}}H_{\nu}(\xi)\\
	\end{array}
	\right)
\end{equation}

where 
\begin{equation}
\xi=\sqrt{eB}(x+\frac{p_{y}}{eB})
\end{equation}
\begin{equation} 	
C{\nu,s}=\sqrt{\frac{\sqrt{eB}}{2^{\nu}\nu! \sqrt{\pi}}}\frac{1}{\sqrt{E_{\nu,s}^{p}+m_{p}^{*}}}
\end{equation}
and $H_{\nu,s}$ are the Hermite polynomials.

The neutron spinors correspond to the usual plane wave solutions of the Dirac
equation with the energy given by equation (15) for spins parallel and
antiparallel to the magnetic field direction.For electrons the problem was
solved in QED by Schwinger to one loop in magnetic field and the energy for
$p_{z}=0$ is given{\cite{label 11}}.

\begin{equation}
E_{\nu,s}^{e}=\sqrt{m_{e}^{2}+(2\nu+1+s)eB}+\frac{\alpha}{2\pi}\frac{eB}{2m_{e}}S
\end{equation}
This expression however breaks down for large magnetic fields for which the
energy is given by

\begin{equation}
E_{\nu,s}^{e}=\sqrt{m_{e}^{2}+(2\nu+1+s)eB}-\frac{\alpha m_{e}}{4\pi}(\log{\frac{2eB}{m_{e}^{\nu}}})^{2} S
\end{equation}
we thus see that for fields strengths of subsequent interest,term proportional
to alpha is negligible and the electron energy is simply taken to be
 
\begin{equation}
E_{\nu,s}^{e}=\sqrt{m_{e}^2+(2\nu+1+s)eB}
\end{equation}

The mean field values $<\sigma_{0}>$,$<\omega_{0}>$ and $<p_{0}^{3}>$ are 
determined by minimizing  the energy at fixed baryon density $n_{B}=n_{p}+n_{n}$ or by 
maximizing the pressure at fixed baryon chemical potential$\mu_{B}$.
We thus get

\begin{equation}
<\sigma_{0}>=\frac{g_{\sigma}}{m_{\sigma}^{2}}(n_{p}^{s}+n_{n}^{s})
\end{equation}
\begin{equation}
<\omega_{0}>={\frac{g_{\omega}}{m_{\omega}^2}}(n_{n}+n_{p})
\end{equation}
\begin{equation}
<p_{0}^{3}>=\frac{1}{2}\frac{g_{\rho}}{m_{\rho}^{2}}(n_{p}-n_{n})
\end{equation}
where $n_{i}$ and $n_{i}^{s}$  the number and scalar number densities 
respectively are given by
\begin{equation}
<n_i>=<\psi_{i}^{\dagger}\psi_{i}> = \frac{1}{(2\pi)^3}
\int \frac{1}{e^{(E_{i}-\mu_{i})\beta}+1} d^3p
\end{equation}
and
\begin{equation}
<n_{i}^{s}> = <\bar{\psi_{i}}\psi_{i}>=\frac{1}{(2\pi)^3} 
\int \frac{m_{i}^{*}}{E_{i}}\frac{1}{e^{(E_{i}-\mu_{i})\beta}+1} \,d^3p
\end{equation}
for each spin state.
In the present case since the energy is different for different spin states,
we need to calculate the densities for spin parallel and antiparallel states
separately.
Using the energy expressions from (14) and (13) carrying out integrals at
$T=0$ we get

\begin{eqnarray}
n_{n,s}
&=&
\frac{1}{2\pi^{2}}[\frac{1}{3}(\mu_{n}^{*2}-(m_{n}^{*}-s\kappa_{n}B)^2)^{\frac{3}{2}}-\frac{s}{2}\mu_{n}^{*2}\kappa_{n}B(s_{m}^{-1}\frac{m_{n}^{*}-s\kappa_{n}}B{\mu^{*}}) \nonumber  \\
&&{} -\frac{\pi}{2}+\frac{m_{n}^{*}-s\kappa_{n}}{\mu_{n}^{*2}}\sqrt{(\mu_{n}^{*2}-(m_{n}^{*}-s\kappa_{n}B)^2)}]
\end{eqnarray}
and
\begin{equation}
n_{p,s}=\frac{eB}{2\pi^2}\sum_{\nu=0}^{\nu_{max}}\sqrt{\mu_{p}^{*2}-m_{p}^{*2}-2eB\nu-\kappa_{p}^2 B^2+2s\kappa_{p} B\sqrt{m_{p}^{*2}+2eB\nu}}
\end{equation}

where

\begin{equation}
\mu_{n}^{*}=\mu_{n}-U_{0}^{n}
\end{equation}
\begin{equation}
\mu_{p}^{*}=\mu_{p}-U_{0}^{p}
\end{equation}
and
\begin{equation}
\nu_{max}=Int(\frac{(\mu_{p}^{*}+s\kappa_{p}B)^2-m_{p}^{*2}}{2eB})
\end{equation}
In the limit of vanishing anomalous magnetic moments,we recover the usual
relations{\cite{label 3}}.
In the region of validity of defining anomalous magnetics moments for protons
and neutrons and for fields of interest ,we can make an expansion in powers of 
$\frac{\kappa_{i}B}{m_{i}^*}$ and can evaluate the densities for highly
degenerate nuclear matter at finite temperatures in powers of
$\frac{\mu^*}{T}$.We get

\begin{equation}
n_{p,s}=\frac{eB}{2\pi^2}\sum_{\nu=0  for  n_{p,+},\nu=1  for  n_{p,-}} p_{F}(p,s)[1-\frac{\pi^2 T^2}{6}\frac{\bar{m_{p}}^2}{p_{F}^{4}(p,s)}+....]
\end{equation}

\begin{equation}
n_{n,s}=\frac{1}{6\pi^2}[p_{F}^{3}(n,s)+\frac{\pi^2 T^2}{2}\frac{(p_{F}^{2}(n,s)+\bar{\mu_{n,s}}^2)}{P_{F}(n,s)}+......]
\end{equation}

\begin{equation}
n_{e}=\frac{eB}{2\pi^2}\sum_{\nu=0}^{\nu_{max}} (2-\delta_{\nu,0})p_{F}(e) (1-\frac{\pi^2 T^2}{6}\frac{\bar{m_{e}}^2}{p_{F}^{4}(e)})
\end{equation}

The corresponding expressions for the scalar densities are given by

\begin{equation}
n_{p,\nu}^{s}= \frac{eB}{2\pi^2} m_{p}^{*} \sum_{\nu} [\ln(\frac{\bar{\mu_{p,s}}+p_{F}(p,s)}{\bar{m_{p}}})-\frac{\pi^2 T^2}{6} \frac{\bar{\mu_{p,s}}}{p_{F}^{3}(p,s)}+....]
\end{equation}

and

\begin{equation}
n_{n,s}^{s}= \frac{m_{n}^{*}}{4\pi^2}[ \bar{\mu_{n,s}} p_{F}(n,s)-m_{n}^{*2}
		\ln(\frac{\bar{\mu_{n,s}}+p_{F}(n,s)}{m_{n}^{*}})+
		\frac{\pi^2 T^2}{3} \frac{\bar{\mu_{n,s}}}{P_{F}(n,s)}+....]
\end{equation}

where 

\begin{equation}
\bar{m_{p}^2}=m_{p}^{*2}+2\nu eB
\end{equation}

\begin{equation}
\bar{m_{e}^2}=m_{e}^2+2\nu eB
\end{equation}

\begin{equation}
\bar{\mu_{p,s}}=\mu_{p}^{*}+s\kappa_{p} B
\end{equation}

\begin{equation}
\bar{\mu_{n,s}}=\mu_{n}^{*}+s\kappa_{n} B
\end{equation}

\begin{eqnarray}
p_{F}(p,s)= \sqrt{\bar{\mu_{p,s}}^2-\bar{m_{p}}^2}\nonumber \\
p_{F}(n,s)= \sqrt{\bar{\mu_{n,s}}^2-m_{n}^{*2}}\nonumber \\
p_{F}(e)= \sqrt{\mu_{e}^2-\bar{m_{e}}^2}
\end{eqnarray}

The thermodynamic potential of the system can now be written as

\begin{eqnarray}
\Omega=
     	&&{}\frac{-g_{\sigma}^2 n_{B}^2}{2 m_{\sigma}^2} 
	-\frac{g_{\rho}^2}{8m_{\rho}^2} (n_{p}-n_{n})^2 
	+\frac{g_{\sigma}^2}{2m_{\sigma}^2}(n_{p}^s+n_{n}^s)^2	
	-\frac{1}{8\pi^2}\sum_{s} [\frac{1}{3} \bar{\mu_{n,s}} 
	p_{F}^{3}(n,s)\nonumber  \\	
	&&{}-\frac{1}{2} m_{n}^{*2}\bar{\mu_{n,s}}p_{F}(n,s)	
	+\frac{1}{2} m_{n}^{*4}\ln(\frac{\bar{\mu_{n,s}}+p_{F}(n,s)}{m_{n}^		{*}})+\frac{2}{3}\pi^2 T^2 \bar{\mu_{n,s}} p_{F}(n,s)+....]\nonumber \\
	&&{} - \frac{eB}{8\pi^2} \sum_{s} \sum_{\nu} [\bar{\mu_{p,s}}p_{F}(p,s)
	-\bar{m_{p}}^2\ln(\frac{\bar{\mu_{p,s}}+p_{F}(p,s)}{\bar{m_{p}}})
	+\frac{\pi^2 T^2}{3}\frac{\bar{\mu_{p,s}}^3}{p_{F}(p,s)}+....]\nonumber  \\
	&&{} - \frac{eB}{8\pi^2} \sum_{\nu} (2-\delta_{\nu,0})[\mu_{e} p_{F}(e)
	-\bar{m_{e}}^2\ln(\frac{\mu_{e}+p_{F}(e)}{\bar{m_{e}}}) 
	+\frac{\pi^2 T^2}{3}\frac{\mu_{e}}{p_{F}(e)}+....]\nonumber  \\
\end{eqnarray}

and the other thermodynamic quantities can be determined by the usual 
thermodynamic relations.In the above 
\begin{equation}
n_{B}=n_{n}+n_{p}
\end{equation}

\begin{equation}
n_{i}=n_{i,+}+n_{i,-}
\end{equation}

In the neutron stars the n-p-e matter is charge neutral and is in $\beta$
equilibrium giving

\begin{equation}
\mu_{n}=\mu_{p}+\mu_{e}
\end{equation}

\begin{equation}
n_{p}=n_{e}
\end{equation}
Using these two equations,$m^*$ is calculated self consistently for a given
baryon density $n_{B}$ and the magnetic field B. The thermodynamic
quantities can then be calculated.
		As a consequence of charge neutrality the magnetic field is
strong enough to quantise the electron motion,the motion of the proton also gets quantized and  the proton fraction increases.
The value of the quantising magnetic field increases with baryon density and
for magnetic field 
$\frac{B}{B_{e}^{c}}\geq \frac{1}{2}(\frac{\mu_{e}}{m_{e}})^2$
where $B_{e}^{c}=m_{e}^2=4 X 10^{13} Gauss$,and for $\mu_{e}$ corresponding 
to given $n_{B}$,the electrons and protons would be in the lowest Landau level leading to interesting consequences.       
\end{section}		

\begin{section}{Weak rates and neutrino emmisivity}

\noindent  As discussed in the introduction direct URCA process
{\cite{ label 1}} the absence of magnetic field can proceed only
above the threshold
of $11 \%$ for the proton fraction.Since the effect of the magnetic
field is to increase 
the 
proton fraction at the same time leading to a fall in the proton's fermi 
momentum,the neutron to proton conversion is enhanced in the presence of 
magnetic field and we would see that in the presence of quantising magnetic 
field,there is no restriction on the proton fraction for the process to 
proceed.We will first consider the effect of magnetic field on the direct 
URCA,modified URCA and pion mediated processes  for the case when the field does
not exceed the ]critical value and will go over to the more interesting case 
where electrons are confined in the lowest Landau level.In the later 
situation,direct URCA and the neutronisation process are the important ones.

\subsection{Weak magnetic field}

\noindent We first consider the effect on the direct URCA process 
$n\rightarrow p+e^{-}+\bar{\nu_{e}}$ and $p+e\rightarrow n+\nu_{e}$,
when the magnetic field is not strong enough to force the electrons in the 
lowest Landau level.Previous calculations{\cite {label 2},\cite{label
4},\cite{label 5}} show that the matrix 
element for the process remains essentially unaffected and the modification 
comes mainly from the phase space factor.In this situation nuclear energies are essentially independent of spin states and we can sum over nuclear spins. 
Treating the nucleons 
non-relativistically and electrons ultra-relativistically,the matrix element squared and summed over spins is given by

\begin{equation}
\sum |M|^2 = 8 G_{F}^2 \cos^2\theta_{c} (4m_{n}^{*} m_{p}^{*}) E_{e} E_{\nu}
		[(1+3g_{A}^2)+(1-g_{A}^2)\cos\theta_{c}]
\end{equation}

where $g_{A}=1.261$ is the axial-vector coupling constant.

The emissivity expression is given by
\begin{equation}
\dot{\cal E}_{\nu} =[\prod_{i} \int \frac{1}{(2\pi)^3 2E_{i}} \,d^3p_{i}] E_{\nu}			\sum\nolimits|M|^2 (2\pi)^4 \delta^{4}(P_{f}-P_{i}) S  
\end{equation}
where the phase space integrals are to be calculated over all particle states
The statistical distribution function $ S=f_{n}(1-f_{p})(1-f_{e})$,where $f_{i}'s$ are the Fermi-Dirac distributions.
We can now evaluate the emissivity in the limit of extreme degeneracy,a 
situation appropriate in neutron star cores by replacing the electron phase 
space factor
\begin{equation}
\int \frac{1}{(2\pi)^3}\longrightarrow \frac{eB}{(2\pi)^2} \sum_{\nu=0}^{\nu_{max}} (2-\delta_{\nu,0}) \int \,dp_{z}
\end{equation}
and using the standard techniques to perform the phase space integrals
{\cite{label 11}} and get

\begin{equation}
\dot{\cal E}=\frac{457\pi}{40320} G_{F}^{2}\cos^2\theta_{c} (1+3g_{A}^{2}) m_{n}^{*}m_{p}^{*} eB T^6 \sum_{\nu=0}^{\nu_{max}} (2-\delta_{\nu,0})\frac{1}
{\sqrt{\mu_{e}^2-m_{e}^2-2\nu eB}}
\end{equation}

where

$$
\nu_{\max}=Int(\frac{\mu_{e}^2-m_{e}^2}{2eB})
$$

In the limit of vanishing magnetic field,the sum can be replaced by an integer
and we recover usual expression{\cite{label 9}}

\begin{equation}
\dot{\cal E}_{\nu}(B=0)=\frac{457\pi}{20160}G_{F}^2 \cos^2\theta_{c} (1+3g_{A}^2) m_{n}^{*}m_{p}^{*} \mu_{e} T^6
\end{equation}

Modified URCA process (2) considered to be the dominant processes for neutron 
star cooling,had been calculated by Frinan and Maxwell {\cite{label 12}} by 
treating the long range NN interactions through one pion exchange potential 
and the short range interactions in the frame work of Landau Fermi liquid.The 
matrix element squared over spins for the process
\begin{equation}
n+n\longrightarrow n+p+e+\bar{\nu_{e}}
\end{equation}
 has been calculated to be

\begin{equation}
\sum\nolimits |M|^2=256 G_{F}^2\cos^2\theta_{c} (16 m_{n}^{*3}m_{p}^{*})g_{A}^{2}(\frac{f}{m_{\pi}})^{4} \alpha_{URCA}\frac{E_{e}E_{\nu}}{(E_{e}+E_{\nu})^2}
\end{equation}

where $f$ is the $\pi-N$ coupling constant $(f^2 \simeq 1)$ and $\alpha_{URCA}$ has 
been estimated to be $=1.54$.
Using the above matrix element square,in the energy loss expression with 
appropriate electron phase space,$\dot{\cal E}_{URCA}$ is calculated to be

\begin{eqnarray}
\dot{\cal E}_{URCA}=
&&{} \frac{11513}{60480}\frac{G_{F}^2\cos^2\theta_{c}}{2\pi}g_{A			}^2 m_{n}^{*3}m_{p}^{*}(\frac{f}{m_{\pi}})^4 
		\alpha_{URCA} T^8	\nonumber  \\ 
&&{} \sum_{\nu=0}^{\nu_{max}}(2-\delta_{\nu,0})(\frac{1}{\sqrt{\mu_{e}^{2}-m_{e}^{2}-2\nu eB}})				\nonumber  \\
\end{eqnarray}

which is the $B\rightarrow 0$ limit goes over to the standard result 
{\cite{label 12}}

\begin{equation}
\dot{\cal E}_{URCA}(B=0)=\frac{11513}{30240}\frac{G_{F}^2\cos^2\theta_{c}}{2\pi}g_{A}^2 m_{n}^{*3}m_{p}(\frac{f}{m\pi})^4 \alpha_{URCA}    \mu_{e}    T^8
\end{equation}

The presence of meson condensates in the core of the neutron stars has been
 considered in the literature { \cite{ label 13}} if present,they could supply the required momentum as
is done by the spectator nucleons in the case of modified URCA process,to beat
the suppression due to energy momentum conservation for degenerate particles
and would result in enhanced cooling through the process
$n+\pi^{-}\longrightarrow n+e^{-}+\bar{\nu_{e}}$ .Maxwell et al 
{\cite{label 14}} constructed the charged pion condensed phase through 
a chiral notation by the unitary operator

\begin{equation}
U(\pi^{c},\mu_{\pi^{-}},k_{c},\theta)= e^{i\int (k_{c}r-\mu_{\pi^{-}}t)V_{3}^0 \,d^3r} e^{i Q^5\theta}
\end{equation}

This generates the charged pion condensed phase with a macroscopic field of 
chemical potential $\mu_{\pi^{-}}$ ,momentum $k_{c}$ and chiral angle 
$\theta$.The participating particles in neutrino emission are the 
quasi-particles which
are the superposition of proton and neutron states.The matrix element squared 
and summed over spins for the quasi-particles $\beta$-decay {\cite{label 3}}
was obtained  in  reference { \cite{ label 14}}and is given by

\begin{eqnarray}
\sum\nolimits |M|^2
&=&
		 8 G_{F}^2\cos^2\theta_{c}(4m_{n}^{*}m_{p}^{*})
		[(1+3\tilde{g_{a}}^2)+(1-\tilde{g_{A}}^2)
		\cos\theta_{e\nu}]                \nonumber   \\
&&{}            E_{e}E_{\nu}\frac{\theta^2}{4} 
		[1+(\frac{g_{a}k_{c}}{\mu_{\pi^{-}}})^2] 
\end{eqnarray}

The energy loss rate in the presence of magnetic field can now be easily 
calculated by assuming that chiral condensates are not really affected in the
presence of external fields of moderate strength { \cite{ label 15}}   and is given by

\begin{eqnarray}
\dot{\cal{E}}_\pi
&=&  \frac{457}{40320} \frac{m_n^{*2}}{k_c}
		\mu_e \frac{eB}{2} T^6 G_F^2 \cos^2 \theta_c
		(1+3 \tilde{g_A}^2)\frac{\theta^2}{4}
		[1+ (\frac{\tilde{g_a}k_c}{\mu_\pi})^2] \nonumber  \\
&&            \sum_\nu (2- \delta_{\nu,0})
		\frac{1}{\sqrt{\mu_e^2 - m_e^2 - 2 \nu eB}} 
\end{eqnarray}
where $\mu_{\pi^{-}}=\mu_{e}$ in equilibrium.In $B\rightarrow 0$ limit ,we get
the standard result { \cite{ label 14}}.
 
During collapse,nucleons are non-degenerate and non-relativistic except at 
the highest densities when neutrinos are trapped and the collapse gets 
underway,neutronization is the most important process for generating 
neutrinos.In this situation the rate for the process 
$e^{-}+p\longrightarrow n+\nu_{e}$ is given by

\begin{equation}
\Gamma=\frac{G_{f}^2\cos^2\theta_{c}}{4\pi^3}(1+3g_{A}^2)n_{p}eB\sum_{\nu} (2-\delta_{\nu,0}) \int_{Q}^{\infty} \frac{E_{e}(E_{e}-Q)^2}{\sqrt{E_{e}^2-m_{e}^2-2\nu eB}} f_{e}(1-f_{\nu}) \,dE_{e}
\end{equation}

where $Q=m_{n}-m_{p}$ and the blocking factor for neutrons has been ignored.
In the initial stage electrons are extremely degenerate and relativistic $E_{e}>>m_{e}$,$\mu_{e}>>kT$ and neutrinos freely escape $f_{\nu}<<1$.

\begin{eqnarray}
\Gamma=
&&{} \frac{G_{F}^2\cos^2\theta_{c}}{4\pi^3}(1+3g_{A}^2)
	n_{p}m_{e}^5\frac{B}{B_{c}}[\sum_{\nu} (2-\delta_{\nu,0})
	\int_{\sqrt{1+\frac{2\nu B}{B_{c}}}}^{\infty}  
	\frac{\varepsilon (\varepsilon-q)^2}{\sqrt{\varepsilon^{2}-1-2\nu
	\frac{B}{B_{c}}}}  f_{e} 
	\,d\varepsilon				\nonumber  \\
&&{} -\sum_{\nu=0}^{\nu_{max}} (2-\delta_{\nu,0}) 
	\int_{\sqrt{1+\frac{2\nu B}{B_{c}}}}^{q}		
	\frac{\varepsilon (\varepsilon-q)^2}
	{\sqrt{\varepsilon^2-1-2\nu\frac{B}{B_{c}}}}  
	f_{e}  \,d\varepsilon]			\nonumber  \\
&&{} \simeq \frac{G_{F}^2\cos^2\theta_{c}}{4\pi^3}(1+3g_{A}^2)n_{p}m_{e}^5 
     \frac{B}{B_{c}} \sum_{\nu} (2-\delta_{\nu,0})  \nonumber \\
&&{}   \int_{0}^{\frac{\mu
      _{e}}{m_{e}}} 
      \frac{\varepsilon^3}{\sqrt{\varepsilon^2-1-2\nu\frac{B}{B_{c}}}} 
        \,d\varepsilon
\end{eqnarray}

where $\varepsilon=\frac{E}{m_{e}}$,$q=\frac{Q}{m_{e}}$ and $\nu_{\max}=Int(\frac{q^2-1}{2\frac{B}{B_{c}}})$ which reduces to 

\begin{equation}
\Gamma(B=0)=\frac{G_{F}^2\cos^2\theta_{c}}{2\pi^3}(1+3g_{A}^2) n_{p}\frac{\mu_{e}^5}{5}
\end{equation}
in the limit of $B\rightarrow 0$.
Later on when neutrinos are trapped $\mu_{e}>>kT$,$\mu_{\nu}>>kT$,but as long as $\mu_{e}>>\mu_{\nu}$,$f_{\nu}=0$ and we get 

\begin{eqnarray}
\Gamma=
&&{} \frac{G_{F}^2\cos^2\theta_{c}}{4\pi^3}(1+3g_{A}^2)
	n_{p}m_{e}^5\frac{B}{B_{c}}[\sum_{\nu} (2-\delta_{\nu,0})
	\int_{\sqrt{1+\frac{2\nu B}{B_{c}}}}^{\infty} \frac{\varepsilon(\varepsilon-q)^2}{\sqrt{\varepsilon^2-1-2\nu
	\frac{B}{B_{c}}}}f_{e} \,d\varepsilon		\nonumber   \\  
&&{}  -\sum_{\nu=0}^{\nu_{max}} (2-\delta_{\nu,0}) \int_{\sqrt{1+\frac{2\nu B}
	{B_{c}}}}^{q+\frac{\mu_{\nu}}{m_{e}}} \frac{\varepsilon(\varepsilon-q)		^2}{\sqrt{\varepsilon^2-1-2\nu\frac{B}{B_{c}}}} 
	f_{e}  \, d\varepsilon]				\nonumber  \\
&&{}  \simeq  \frac{G_{F}^2\cos^2\theta_{c}}{4\pi^3}(1+3g_{A}^2)n_{p}m_{e}^5 
	\frac{B}{B_{c}} \sum_{\nu} (2-\delta_{\nu,0})  \nonumber \\
&&{}	\int_{q+\frac{\mu_{\nu}}{m_{e}}}^{\frac{\mu_{e}}{m_{e}}} 
	\frac{\varepsilon^{3}}
	{\sqrt{\varepsilon^{2}-1-2\nu\frac{B}{B_{c}}}}\,d\varepsilon
\end{eqnarray}

which reduces to
 
\begin{equation}
\Gamma(B=0)=\frac{G_{F}^2\cos^2\theta_{c}}{2\pi^3}(1+3g_{A}^2)n_{p}\frac{\mu_{e}^5-\mu_{\nu}^5}{5}
\end{equation}

in the $B\rightarrow 0$ limit.

\end{section}
\begin{section}{Quantising magnetic field}

\noindent In the case of super strong magnetic fields such that
$2eB>p_{F}(e)^2$ 
all electrons occupy the Landau ground state at T=0 which corresponds to
$\nu=0$ state with electron spins pointing in the direction opposite to the magnetic
field.As has been discussed above,charge neutrality now forces the degenerate
non-relativistic protons also to occupy the lowest Landau level with proton 
spins pointing in the direction of the fiel.In this situation we can no longer consider the matrix elements to be
unchanged and they should be evaluated using the exact solutions of Dirac 
equation.Further because nucleons have anomalous magnetic moment,matrix 
elements need to be evaluated for specific spin states separately.
The electron wave function in $\nu=0$ state has energy 
$E_{e}=\sqrt{m_{e}^2+p_{ez}^2}$ with a wave function

\begin{equation}
\psi_{e}(r)= (\frac{eB}{\pi})^{1/4}
	\frac{1}{\sqrt{L_{y}L_{z}}}e^{-i(E_{e}t-p_{ey}y-p_{ez}z)}
 		e^{\frac{-\xi^2}{2}}U_{e,-1}(E_{e})
\end{equation}

and the positive energy spinor in $\nu=0$ state is given by

\begin{equation}
 U_{e,-1} = \frac{1}{\sqrt{E_{e}+m_{e}}} \left( 
       		\begin{array}{c}	
		0 \\
		E_{e}+m_{e} \\
		0 \\
		-p_{ez} \\
	     	\end{array}
       		\right)
\end{equation}

Protons are treated non-relativistically,from equation(14) we define
\begin{equation}
\tilde{m_{p}}=m_{p}^{*}-\kappa_{p}B
\end{equation}

and the energy in $\nu=0$ state is
\begin{equation}
E_{p}\simeq \tilde{m_{p}}+\frac{P_{z}^2}{2\tilde{m_{p}}}+U_0^{p}
\end{equation}

The proton wave function is given by
\begin{equation}
\Psi_{p}(r)= (\frac{eB}{\pi})^{1/4} \frac{1}{\sqrt{L_y L_z}}
	e^{-i(E_pt - P_y y - P_z z)} e^{-\frac{\xi^2}{2}U_{p,+1}(E_{p})}
\end{equation}
where
\begin{equation}
\xi=\sqrt{eB}(x-\frac{P_{y}}{eB})
\end{equation}

and $U_{p,+1}$ is the non-relativistic spin up operator

\begin{equation}
 U_{p,+1}= \frac{1}{\sqrt{2\tilde{m_{p}}}} \left(
					\begin{array}{c}
					\chi_{+} \\
					0 \\
					\end{array}
					\right)
\end{equation}
For neutrons we have

\begin{equation}
\Psi_{n,s}(r)=\frac{1}{\sqrt{L_{x}L_{y}L_{z}}}e^{-ip_{n,s} \cdot r}U_{n,s}(E_{n,s})
\end{equation}
\begin{equation}
 U_{n,s} = \frac{1}{\sqrt{2m_{n}^{*}}} \left(
				\begin{array}{c}
					\chi_{s} \\
					0 \\
				\end{array}
				\right)
\end{equation}
and
\begin{equation}
E_{n,s} \simeq m_{n}^{*}+\overrightarrow{p_{n}}^2-\kappa_{n} B s+U_0^{n}
\end{equation}

in the non-relativistic limit.The neutrino wave function is given by

\begin{equation}
\Psi_{\nu}(r)=\frac{1}{\sqrt{L_{x}L_{y}L_{z}}}e^{-ip_{\nu\cdot r}}U_{\nu,s}(E_{\nu})
\end{equation}

here $U_{\nu,s}$ is the usual free particle spinor,$\chi_{s}$ is the spin 
spinor and the wave function have been normalised in a volume 
$V=L_{x}L_{y}L_{z}$ and we have used the normalization $\sum \bar{U_{\alpha}}U_{\beta} =2m\delta_{\alpha,\beta}$
We will now calculate the matrix element for the direct URCA process (1) for 
different polarisation states of neutron.The neutrino polarisation will be 
summed over
\begin{equation}
M_s=
\frac{G_F\cos\theta_c}{\sqrt{2}} 
\int\bar{\Psi}_{N,s}(r)\gamma_{\mu}(1-g_{A}\gamma_{5})\Psi_{p,+1}\bar{\Psi_{\nu}}(r)\gamma^{\mu}(1-\gamma_{5})\Psi_{e,-1}(r) d^4r
\end{equation}

Substituting for the wave functions and carrying out the integrals,we obtain

\begin{equation}
M_{s}= K_{s} \bar{U_{n,s}}\gamma_{\mu}(1-g_{A}\gamma_{5})U_{p,+1}\bar{U_{\nu}}
	\gamma^{\mu}(1-\gamma_{5})U_{e,-1}
\end{equation}

where
\begin{equation}
K_{s}=\frac{2\pi^3 G_{F}\cos\theta_{c}}{\sqrt{2}(L_{y}L_{z})^2 L_{x}}\delta(E_{n}-E_{p}-E_{e}-E_{\nu})\delta(p_{nz}-p_{z}-p_{ez}-p_{\nu z})\delta(p_{ny}-p_{y}-p_{ey}-p_{\nu y}) e^{-Q^2}
\end{equation}

and

\begin{equation}
Q^2=\frac{p_{y}^2+p_{ey}^2}{2eB}- \frac{(p_{ey}+P_{y})+i(p_{nx}-p_{\nu x})^2}{4eB}
\end{equation}

now calculate the matrix element squared and summed over neutrino states to get

\begin{equation}
\sum\nolimits |M_{+}^{\dagger}M_{+}|=|K_{+}|^{2}4(4
		m_{n}^{*}\tilde{m_{p}})(1+g_{A})^2(E_{e}+p_{ez})
		(E_{\nu+p_{\nu z}})
\end{equation}
and

\begin{equation}
\sum\nolimits |M_{-}^{\dagger}M_{-}|=|K_{-}|^{2} 16 (4 m_{n}^{*}\tilde{m_{p}})
g_{A}^2(E_{e}+p_{ez})(E_{\nu}-p_{\nu z})
\end{equation}

The neutrino emissivity is now given by

\begin{eqnarray}
\dot{\cal E} =
&&{} \frac{2G_{f}^{2}\cos^2\theta_{c}}{V^2 L_{y}^{2}L_{z}^{2}L_{x}}
     (4m_{n}^{*}\tilde{m_{p}}) 
     \int_{\frac{-eBL_{x}}{2}}^{\frac{eBL_{y}}{2}} \frac{L_{y}}{2\pi}
      \,dp_{ey} \int_{-\infty}^{\infty} \frac{L_{z}}{2\pi 2E_{e}} \,dp_{ez}
     \int_{\frac{-eBL_{x}}{2}}^{\frac{eBL_{x}}{2}} \frac{L_{y}}
     {2\pi} \,dp_{y}                                      \nonumber  \\
&&{} \int_{ - \infty}^{\infty} \frac{L_{z}}{2\pi 2\tilde{m_{p}}}\,dp_{z} 
     \int \frac{V}{(2\pi)^{3} 2m_{n}^{*}} \,d^{3}p_{n}
     \int \frac{V}{(2\pi)^{3} 2E_{\nu}} \,d^{3}p_{\nu}   \nonumber   \\
&&{} E_{\nu} (2\pi)^{3} [(1+g_{A})^2 (E_{\nu}+p_{\nu,z})
     \delta(E_{n}-E_{p}-E_{e}-E_{\nu})
     f_{n,+}                                             \nonumber  \\
&&{} +4g_{A}^2 (E_{\nu}-p_{\nu,z})\delta(E_{n,-}-E_{p}-
     E_{p}-E_{\nu}) f_{n,-}](E_{e}+p_{ez})(1-f_{e})(1-f_{p})      \nonumber \\ 
&&{} \delta(p_{n,z}-p_{z}-p_{ez}-p_{\nu z})              
     \delta(p_{ny}-p_{y}-
     p_{ey}-p_{\nu y})                                  \nonumber \\
&&{} e^{\frac{-1}{2eB}(p_{\nu x}+p_{nx})^2+
    (p_{y}+p_{ez})^2}                              
\end{eqnarray}

Integrals over $dp_{y}$ and $dp_{ey}$ can be performed by using the 
y-component momentum delta function.The integrals over $dp_{ez}$ and $dp_{z}$
are converted into integrals over $dE_{e}$ and $dE_{p}$ respectively and 
making use of the fact that for strongly degenerate matter,particles at the 
top of their respective fermi seas alone contribute.The integral over 
neutron solid angle is performed by using the z-component momentum conserving
delta function and neglecting neutrino momentum.The integral over angle can be
performedddd explicitly and we obtain

\begin{eqnarray}
\dot{\cal E}=
&&{}	\frac{2 G_{F}^2\cos^2\theta_{c}}{(2\pi)^5} eB 
	\frac{m_{n}^{*}\tilde{m_{p}}}{p_{F}(p)} \int E_{\nu}^3 
	\,dE_{e} \,dE_{p} \,dE_{n} \,dE_{\nu} [(1+g_{A})^2	\nonumber\\ 
&&{}	\delta(E_{n,+}-E_{p}-E_{e}-E_{\nu})f_{n,+}+4g_{A}^2 
	\delta(E_{n,-}-E_{p}-E_{e}-E_{\nu})f_{n,-}]		\nonumber\\
&&{}	[e^{\frac{-1}{2eB}(p_{f}^{2}(n)-4p_{F}^{2}(c))}
	\theta(p_{F}^{2}(n)-4p_{F}^{2}(e))+			\nonumber\\
&&{}	e^{\frac{-1}{2eB}p_{F}^{2}(n)}\theta(p_{F}(n))](1-f_{p})(1-f_{e})
\end{eqnarray}

The two $\theta$ functions correspond to the fact that the z-component momentum coservation $p_{nz}=p_{z}+p_{ez}=p_{F}(p) \pm p_{F}(e)$ depending on whether both electrons and protons in their Landau ground states move in the sane direction or in the opposite directions.Further since charge neutrality implies 
$p_{F}(p)=p_{F}(e)$,$p_{nz}=2p_{F}(e)$ or zero.The energy integrals can now be performed by the standard techniques for degenerate matter and we get

\begin{eqnarray}
\dot{\cal E}=
&&{}	\frac{457\pi}{40320} G_{F}^2\cos^2\theta_{c} eB 
	\frac{\tilde{m_{p}}m_{n}^{*}}{p_{F}(e)} T^6 
	[\frac{(1+g_{A})^2}{2}					\nonumber\\
&&{}	{\theta(p_{F}^{2}(n,+)) e^{\frac{-p_{F}^{2}(n,+)}{2eB}}+
	\theta(p_{F}^{2}(n,+)-4p_{F}^{2}(e)) 
	e^{\frac{-(p_{F}^{2}(n,+)-4p_{F}^{2}(e))}{2eB}}}	\nonumber\\
&&{}	+ 2g_{A}^{2}{p_{F}(n,+)\longrightarrow p_{F}(n,-)}]	\nonumber\\
\end{eqnarray}

where $p_{f}(n,+)$ and $p_{F}(n,-)$ are the neutron fermi momenta for the neutron spins along and opposite to the magnetic field direction respectively and are given by (sec. eq.(60))

\begin{equation}
\frac{p_{F}^{2}(n,\pm)}{2m_{n}^{*}}=\mu_{n}-m_{n}^{*}-U_{0}^{n} \pm \frac{\kappa_{n}B}{2m_{n}^{*}}
\end{equation}

The energy loss rate due to $p+e^{-}\rightarrow n+\nu_{e}$ gives the same contribution under equilibrium and the total emissivity is thus a factor of two greater than (69)
We thus see as advertised that in the presence of quantising magnetic field the inequality $p_{F}(e)+p_{F}(p)\geq p_{f}(n)$ is no longer required to be satisfied for the process to proceed.Regardless of the value of proton fraction determined by $p_{F}(p)=p_{F}(e)$,we get non zero energy loss rate.The rate however,
would show an increase at the threshold of the proton fraction 
$Y_{p} \geq 11\%$.

During collapse when the neutronization process is important,the density is of the order of 
$10^{12} gm/cc $and the nucleons are non-degenerate and 
non-relativistic.The strong interaction effects can be ignored.The nucleon expressions can be simplified and we have from equations(68)and (74)

\begin{equation}
E_{p} \simeq m_{p}-\kappa_{p}B+\frac{p_{z}^2}{2\tilde{m_{p}}}	\nonumber\\
\end{equation}

\begin{equation}
E_{n,s} \simeq m_{n}-\kappa_{n}B+\frac{\overrightarrow p_{n}^{2}}{2m_{n}} \nonumber\\
\end{equation}
 
defining $Q_{s}=E_{n,s}-E_{p} \simeq Q-(\kappa_{n}s-\kappa_{p})B $,the rate for
the neutronization process can be similarly evaluated in the limit 
$E_{e}\sim \mu_{c}>>Q_{s}$ and we obtain 

\begin{eqnarray}
\Gamma \simeq
&&{}	\frac{2 \sqrt{2}G_{F}^{2}\cos^2\theta }{3\pi^3}(m_{p})^{\frac{3}{2}}
 n_{p} 
	\int \,dE_{e} E_{e}^{\frac{5}{2}}\Bigg[(1+g_{A})^2 
	e^{\frac{E_{e}Q_{+}}{eB}}-		\nonumber\\
&&{}	2g_{A}^2 \frac{Q_{-}}{E_{e}} e^{\frac{E_{e}Q_{-}}{eB}}\Bigg]f_{e}(E_e)
\end{eqnarray}

\end{section}

\begin{section}{Results and Discussions}
We now present our results.For the case of non-interacting,cold,degenerate 
matter,containing nucleons and electrons with neutrinos freely streaming out,
we choose $n_{B}$ to lie in the range of ($\sim 0.5$ to $6$)$n_{0}$,a 
situation appropriate for the neutron star core and show the composition of 
matter as a function of density in figure 1 for different values of the 
magnetic field.For $B=0$,the proton fraction remains less than the threshold 
value 
$(\frac{n_{p}}{n_{B}}<11\%)$ for direct URCA process to take place,even for 
$n_{B}$ as large as $10 n_{0}$.With the increase in magnetic field,the proton fraction rises steadily and crosses the threshold for $n_{B}<6 n_{0}$ for 
$B=10^{5} (MeV)^2$ say.A further increase in B results in the presence of 
more protons than neutrons for example,for $B=5 X 10^{5}(MeV)^2$ there are 
more protons than neutrons upto $n_{B} \leq 3 n_{0}$ and at large 
densities,the proton fraction again decreases but remains large.
Any further increase in B 
may result in a prominently proton matter star.The effect of including 
anomalous magnetic moment of nucleons is to increase the proton fraction at 
all densities as can be seen from figure 1 and the proton fraction rises 
above the neutron fraction.Effect of including interactions (fig.2) is to 
raise the proton fraction and the threshold for direct URCA process comes 
down to $n_{B}\geq 1.5n_{0}$.
Magnetic field has the effect of generally raising the proton fraction and 
for the highest magnetic field $B\simeq 5x10^{5} MeV^{2}$ considered here,
there are far  more protons than neutrons in the star.
The effect of magnetic field on neutrino emissivity is shown in table 1 
where the ratio of the emissivity 
$(R=\frac{{\cal E}_{\nu}(B)}{{\cal E}_{\nu}(0)})$ 
for the direct URCA process is shown as a function of density.These results 
are for the weak magnetic field which is our case of degenerate matter  
could even be as large as $10^{4} MeV^{2}$.The interesting case of magnetic 
field that is capable of totally polarising the electrons and protons is 
shown in figure 3 where we have plotted the neutrino emissivity in units of
${\cal E}_{0}=\frac{457\pi}{20160}G_{F}^{2}\cos^2\theta_{c} T^{6} m_{p}^{3}$.
As has been discussed above,in this case the threshold for direct URCA process is evaded and the emissivity is enhanced by upto two orders of magnitude.

\[ {\begin{array}{cccc}
\hline
 	 &	B=10^{2} &	B=10^{3} 	&	B=10^{4} \\
\hline
n_{B}	 &   	R 	 &	R 		&	R 	 \\
\hline
0.23 	 &	0.94 	&	0.94 		&	0.65 \\
0.36 	 &	0.96 	&	0.89 		&	0.62 \\
0.54 	 &	0.98 	&	1.23 		&	1.10 \\
0.71 	 &	0.99 	&	1.29 		&	0.89 \\
0.98 	 &	0.97 	&	1.03 		&	1.06 \\
\hline
\end{array}} \]  
\vskip 0.2cm
Table 1.Ratio $R=\frac{{\cal E}_{\nu}(B)}{{\cal E}_{\nu}(0)}$as a function 
of density for different values of magnetic field $(in  MeV^{2})$
\vskip 0.5cm

Similar situation obtains for neutronization rate during collapse   when the nucleons are non-degenerate and density $\sim 10^{12}gm/cc$.In this case even the magnetic field of moderate strength $(B>10^{2}MeV^{2})$ is able to polarise 
the electrons completely,but unlike the cold matter,the charge neutrality here
does not force the protons to be totally polarised.However for 
$B\sim 10^{4}MeV^{2}$ and above,the protons too are completely polarised.
In figure 4 we have plotted the reaction rates $\Gamma$ in units of 
$\Gamma_{0}=\frac{10^{6}G_{F}^{2}\cos^2\theta_{c} m_{e}^{8}}{\pi^{3}}$ for $B=0$ 
and $B=10^{4}MeV^{2}$for non-interacting non-degenerate hot nuclear matter
at $T=5 MeV$ as a function of density and find the rates to be 
greatly enhanced in the presence of strong magnetic field.
Strong magnetic field thus changes the composition of nuclear matter in a very
substantial way and enhances the cooling rates as well as the neutronizaton
rates and this would have serious implications for neutron and pulsar dynamics.
\end{section}
 
\pagebreak

\bibliography{plain}

\pagebreak

%
\begin{figure}
\epsfig{file=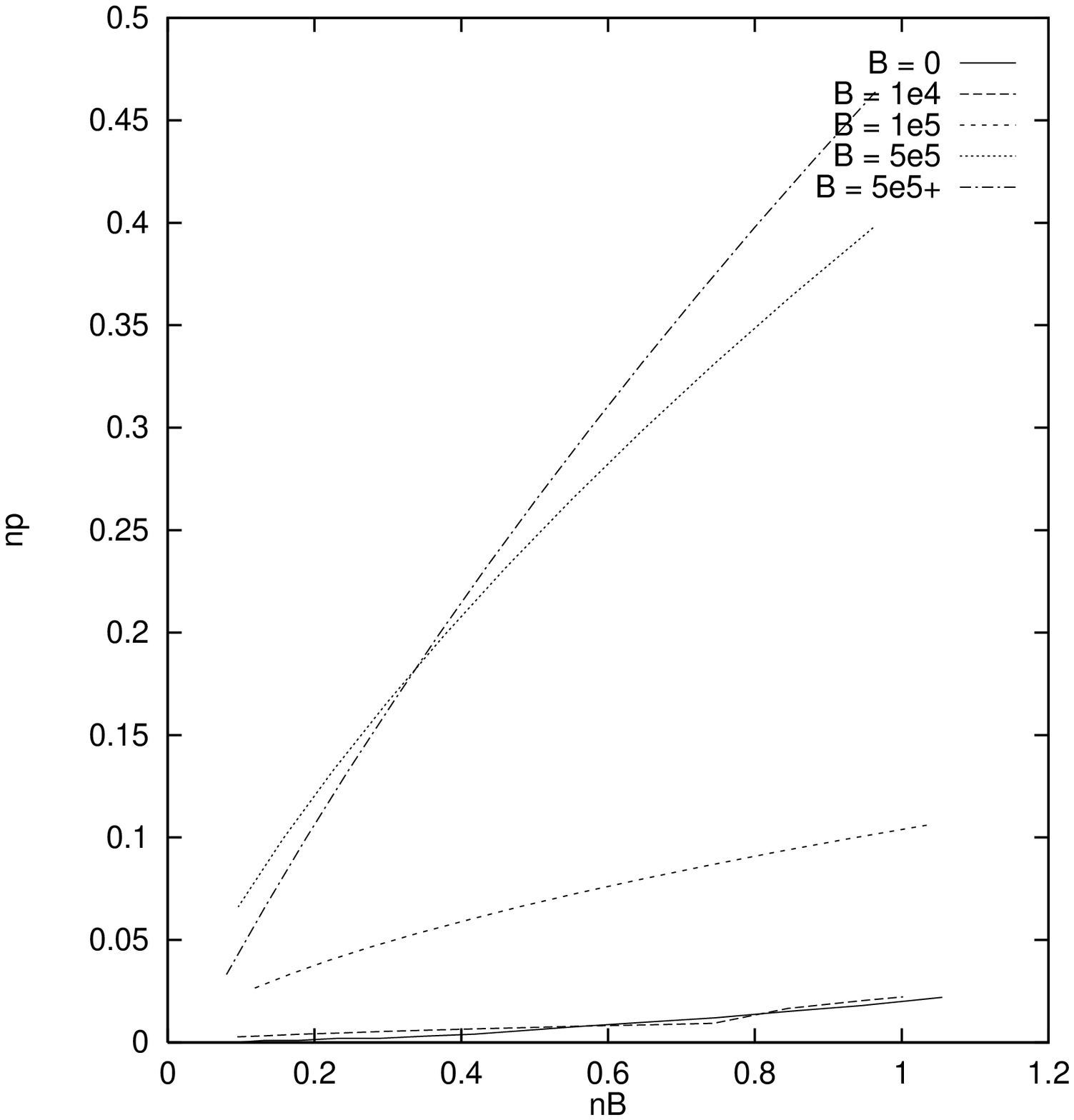,width=10cm,height=5cm}
\end{figure}
\begin{figure}
\epsfig{file=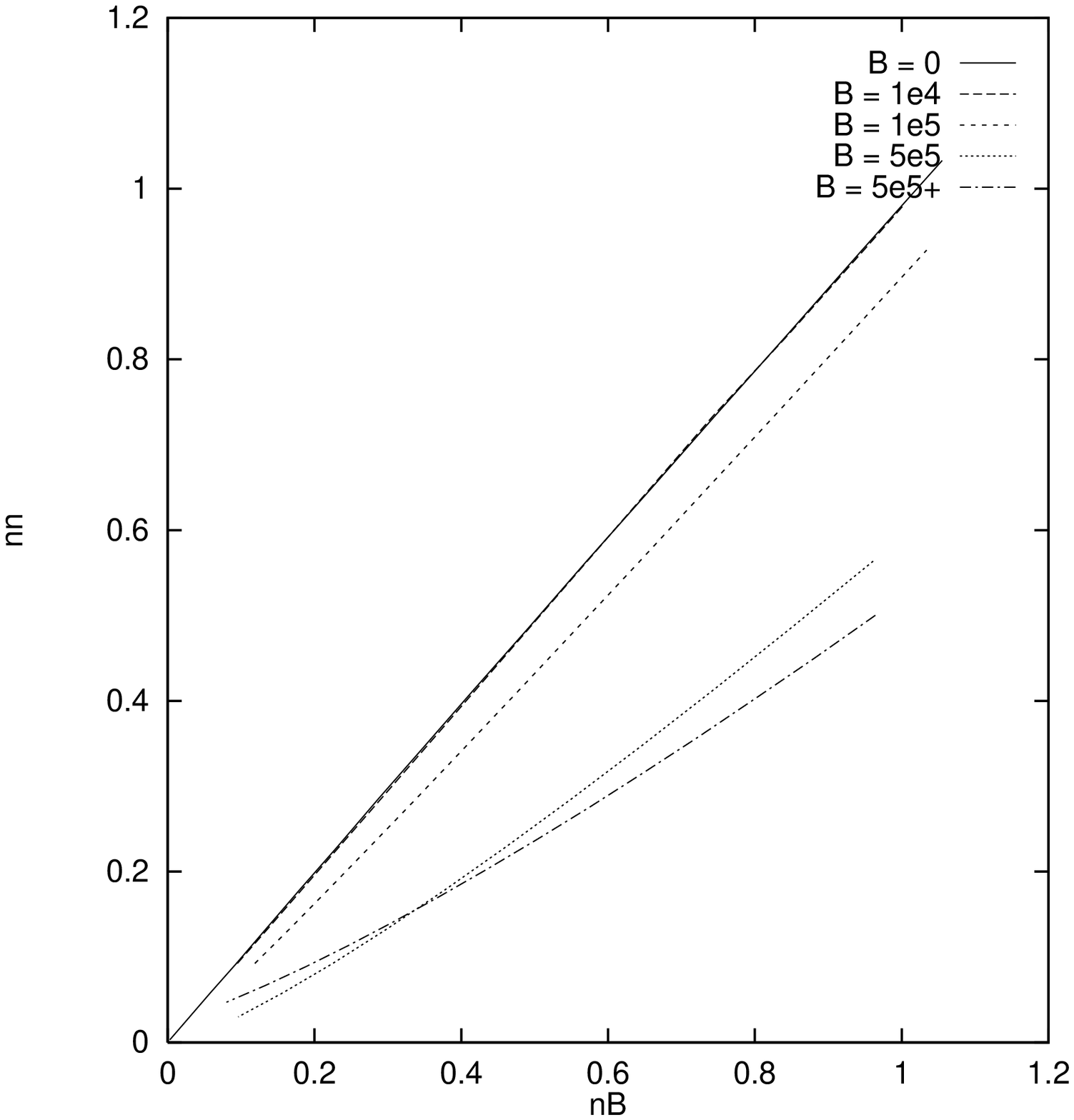,width=10cm,height=5cm}
\caption{Effect of magnetic field on the composition of non-interacting 
cold nuclear matter.The baryon density,neutron density and proton density 
are given in units of $fm^{-3}$.The effect of including anomalous magnetic 
moment of nucleons on the composition is shown for $B=5X10^{5}$ by the 
dashed-dotted curve.The magnetic field is in units of $MeV^{2}$.}
\end{figure}
\vfill
\eject
\begin{figure}
\vskip 2truecm
\epsfig{file=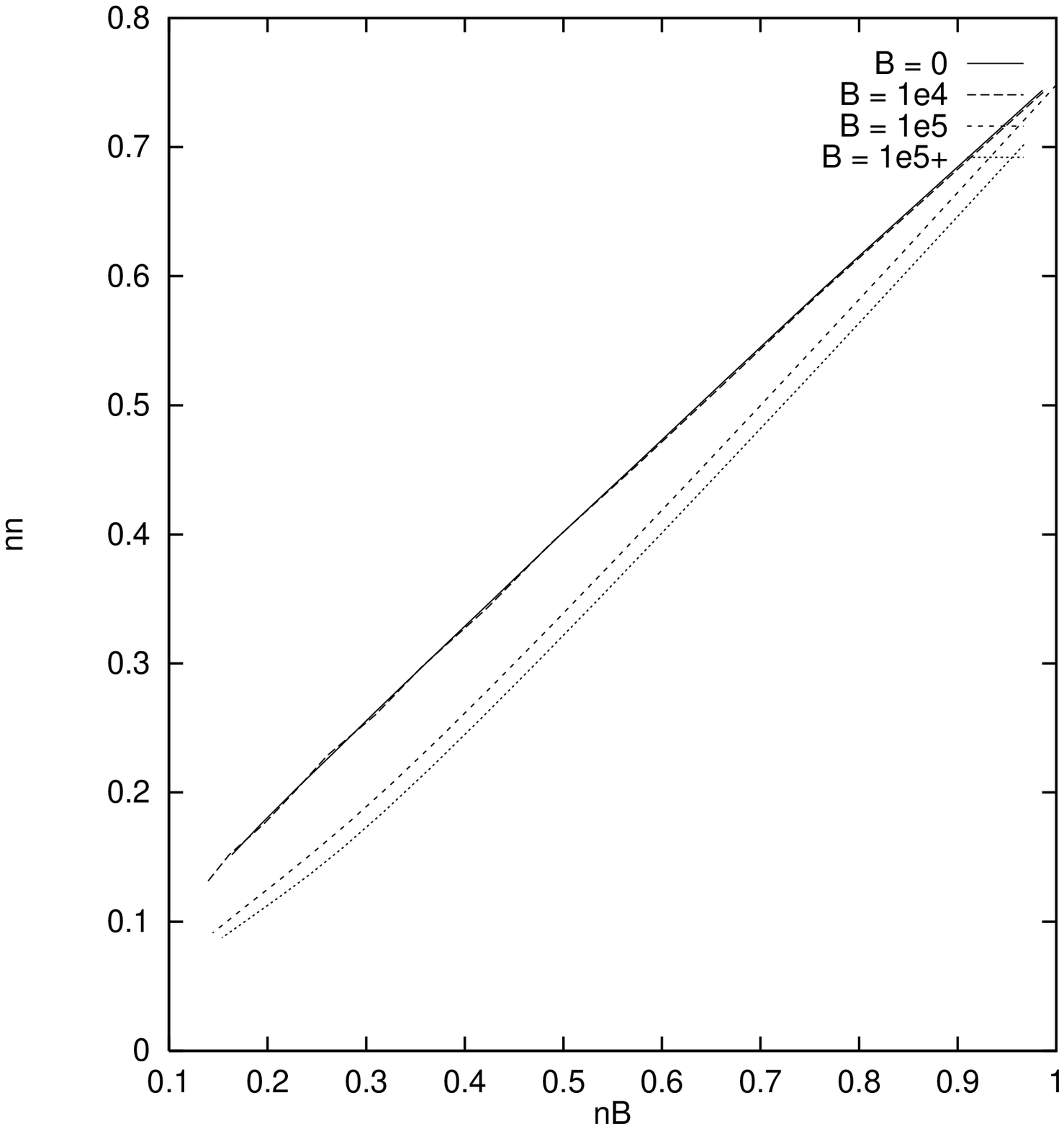,width=10cm,height=5cm}
\end{figure}
\begin{figure}
\epsfig{file=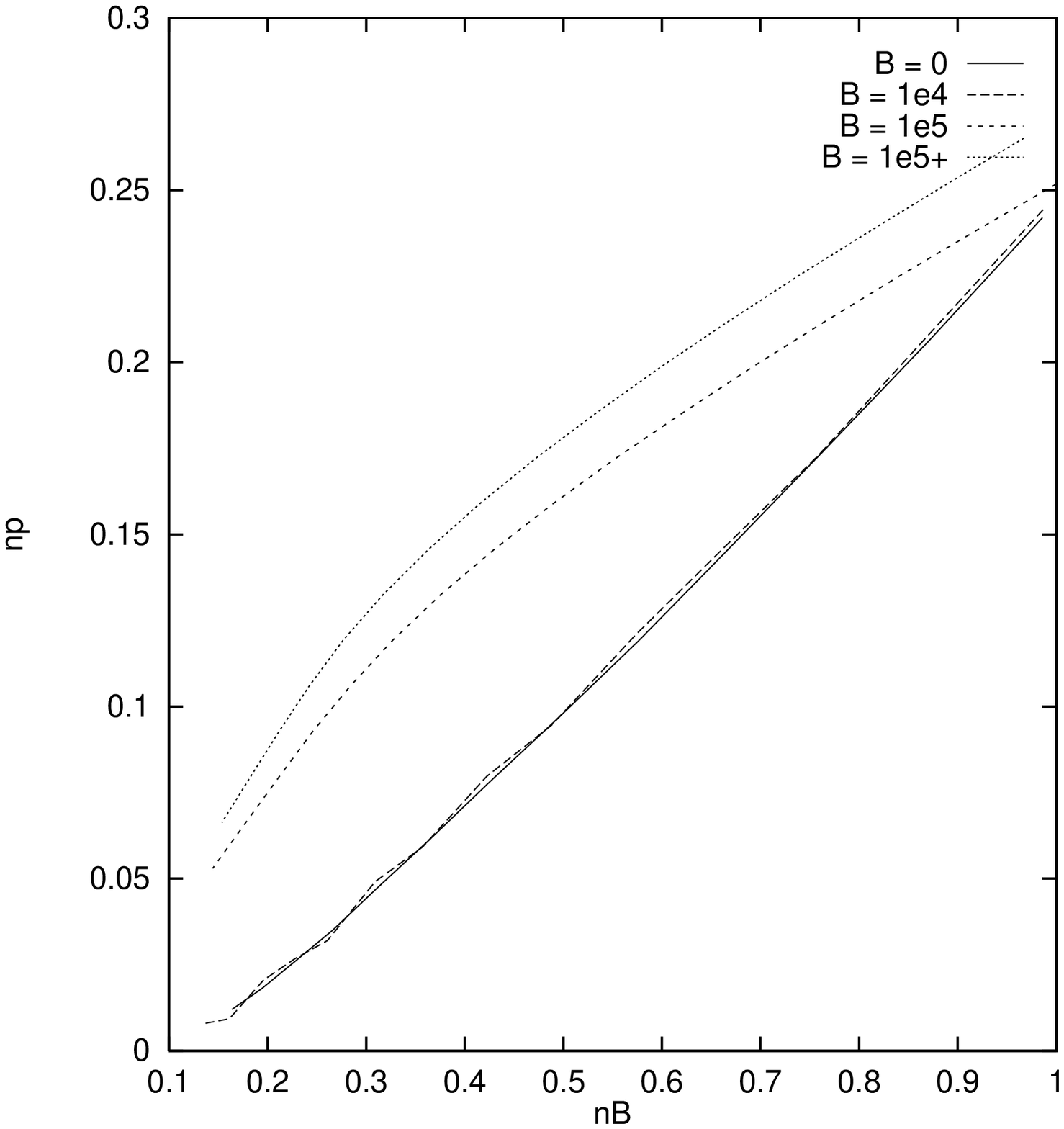,width=10cm,height=5cm}
\end{figure}
\begin{figure}
\epsfig{file=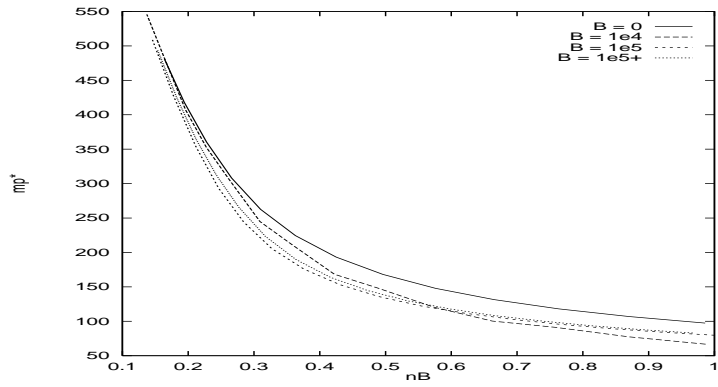,width=10cm,height=5cm}
\caption{Effect of magnetic field on the composition and effective proton 
mass as a function of baryon density for cold interacting nuclear 
matter for 
different values of the magnetic field.The densities are in units of 
$fm^{-3}$,proton effective mass in Mev and magnetic field in units of
$MeV^{2}$.The effect of including anomalous magnetic moment of nucleons on 
the composition and on proton mass is shown for $B=10^{5}$ by the 
dotted curve.}
\end{figure}
\begin{figure}
\epsfig{file=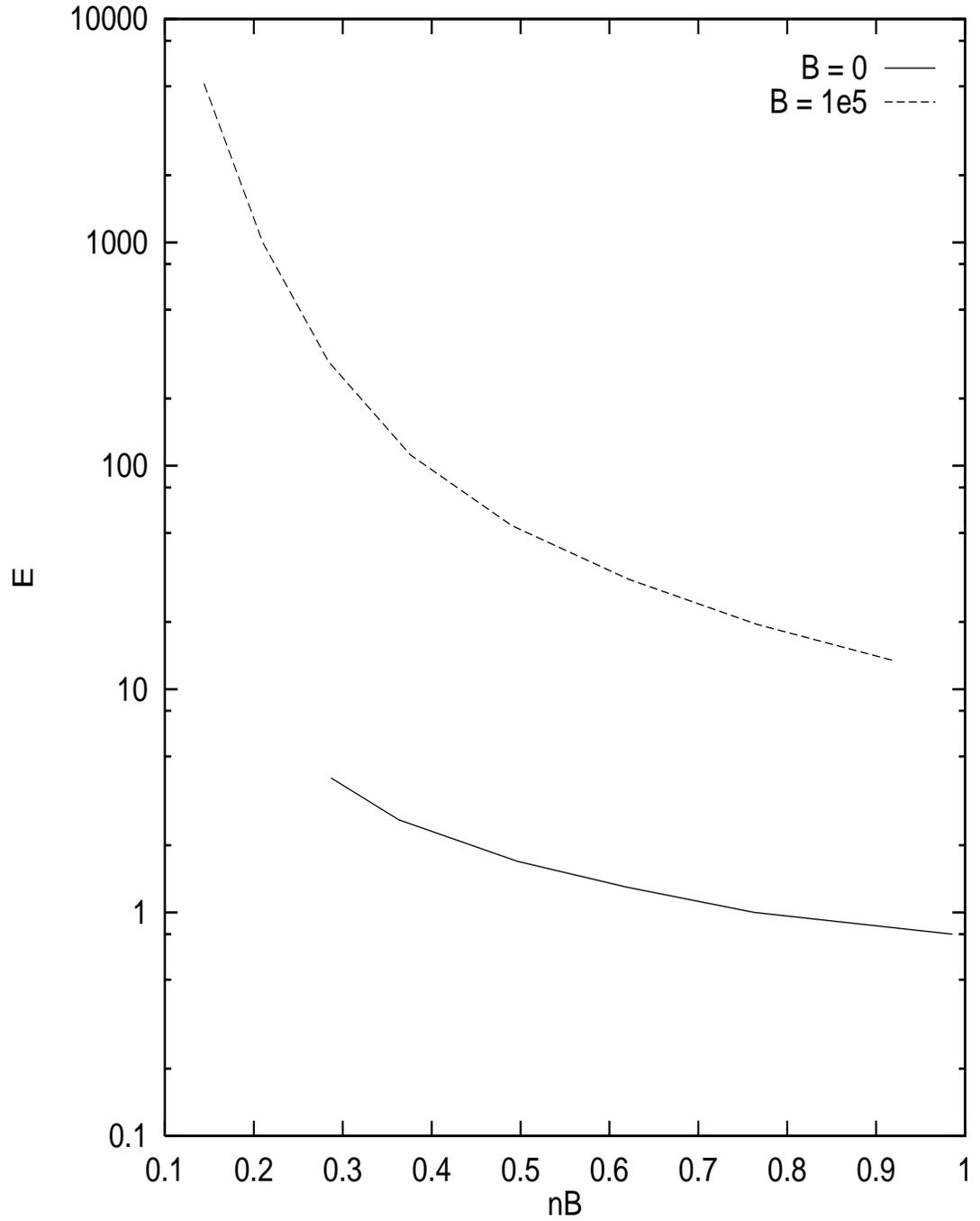,width=15cm,height=18cm}
\caption{Neutrino emissivity in units of${\cal E}_{0}$(see text) due to direct URCA process as a function of baryon density for $B=0$ and $B=10^{5}
MeV^{2}$
when the electrons and protons are completely polarised.}
\end{figure}
\begin{figure}
\epsfig{file=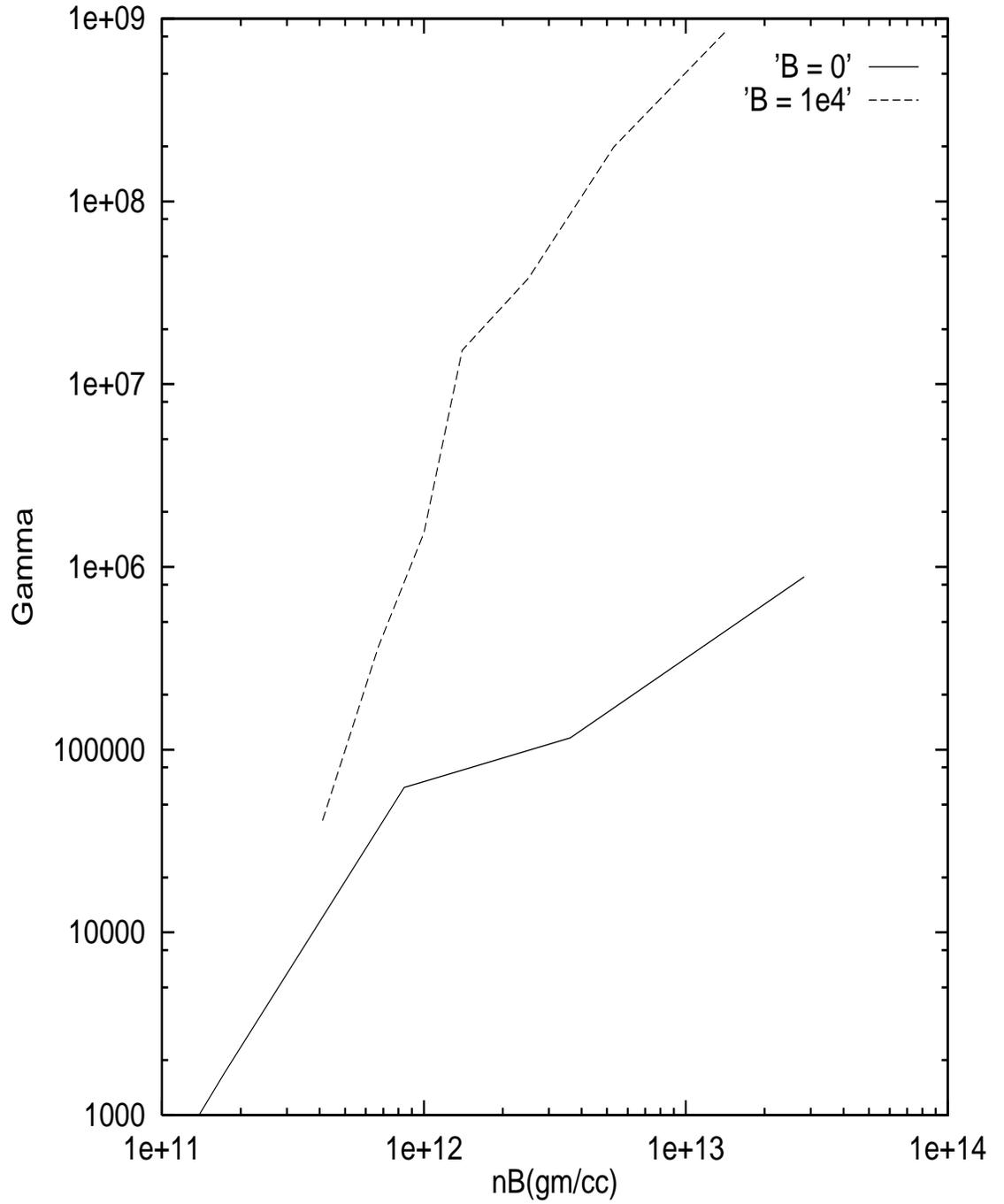,width=15cm,height=18cm}
\caption{Neutronization rate in units of $\Gamma_{0}$(see text) as 
a function 
of baryon density for $B=0$ and $B=10^{4}MeV^{2}$ for hot non-interacting 
non-degenerate nuclear matter at $T=5 MeV$.The baryon density is given in 
gm/cc.}
\end{figure}

\end{document}